\documentclass[reprint,amsmath,amssymb,aps,prb,showpacs,twocolumn]{revtex4}
\usepackage[utf8]{inputenc}
\usepackage{graphicx}
\usepackage{dcolumn}
\usepackage{bm}
\usepackage[breaklinks]{hyperref}
\usepackage{threeparttable}
\usepackage{bbold}
\usepackage{subfigure}
\usepackage{color}
\usepackage[normalem]{ulem}
\usepackage{cancel}
\usepackage{braket}
\usepackage{mathrsfs}

\renewcommand*{\HyperDestNameFilter}[1]{\jobname-#1}

\renewcommand{\vec}[1]{\mathbf{#1}}
\newcommand{\vn}[1]{{\vec{#1}}}
\newcommand{\vht}[1]{{\boldsymbol{#1}}}

\begin{document}

\preprint{APS/123-QED}

\title{Spin-orbit torques in L1$_0$-FePt/Pt thin films driven by electrical and thermal currents}

\author{Guillaume G\'eranton}
\email{g.geranton@fz-juelich.de}
\author{Frank  Freimuth}
\author{Stefan Bl\"ugel}
\author{Yuriy Mokrousov}
\affiliation{Peter Gr\"unberg Institut and Institute for Advanced Simulation,
Forschungszentrum J\"ulich and JARA, 52425 J\"ulich, Germany}

\date{\today}

\begin{abstract}
Using the linear response formalism for the spin-orbit torque (SOT) we compute from first principles 
the SOT in a system of two layers of L1$_0$-FePt(001)
 deposited on an fcc Pt(001) substrate of varying thickness. We find that at room temperature the values of the SOTs that are even and odd with respect to magnetization generally lie in the range of values measured and computed for Co/Pt
 bilayers. We also observe that the even SOT is much more robust with respect to changing 
 the number of layers
 in the substrate, and as a function of energy it follows the general trend of the even SOT
 exerted by the spin Hall current in fcc Pt. The odd torque, on the other hand, is 
strongly affected by modification of the electronic structure for a specific energy window in the limit of very thin films. Moreover, taking the system at hand as an example, we compute the values of the thermal spin-orbit torque (T-SOT).
 We predict that the gradients of temperature which can be experimentally created in this
 type of systems will cause a detectable torque on the magnetization. We also underline
 the correlation between the even T-SOT and the spin Nernst effect, thus motivating a more
 intensive  experimental effort aimed at observation of both phenomena.

\end{abstract}

\maketitle

\section{Introduction}
The possibility of manipulating the magnetization of a ferromagnet by spin-polarized electric currents was first pointed out by Berger\cite{PhysRevB.54.9353} and Slonczewski.\cite{Slonczewski1996L1} After its experimental observation in 1999\cite{Sun1999157} there has been growing interest in making use of this phenomenon in spintronics devices,~e.g.~for 
switching the magnetization in spin valves. Spin-transfer torque random-access memory (STT-MRAM) has the advantages of better scalability and lower power consumption over conventional magnetoresistive random-access memory (MRAM), where the information is written by an external magnetic field. However, the current density recquired to switch the magnetization is preventing STT-MRAM size from scaling down and overcoming this problem is a central aim of current research efforts in the field of spintronics. From the fundamental point of view, it has become necessary to develop an understanding of the interplay between the magnetization and the currents driven by electric fields, as well as temperature gradients.

While spin-transfer torques rely on the exchange of spin angular momentum between two magnets with different direction of the magnetization, the so-called spin-orbit torques (SOTs) have been discovered only recently\cite{Chernyshov:154015,Miron:154509,MihaiMiron:155006} and they are attributed to  the spin-orbit-mediated exchange of angular momentum between the crystal lattice and the magnetization. This type of torques exists also in systems with collinear magnetization when inversion symmetry is broken, and it has been shown that SOTs can lead to a reversal 
of a ferromagnetic magnetization without the help of an additional polarizing layer.\cite{Miron:155001,Liu04052012,PhysRevLett.109.096602} Moreover, SOTs were shown to lead to a very fast domain wall motion in thin films at low current density.\cite{Emori:155277,Miron:154509,Ryu:154510} This suggests that SOTs could play a crucial role in the next generation of spintronics devices.

On the theoretical side, two mechanisms have been proposed that give rise to SOTs in heavy metal/ferromagnetic bilayers. The first one is due to the torque exerted on the magnetization by the spin current 
from the spin Hall effect (SHE) of the heavy metal.
The second one is due to the non-equilibrium spin density that is generated at the interface when the distribution function of the system is driven out of equilibrium by an electric field, which is expected, e.g., in the Rashba model.
While the sign and the amplitude of the SOT due to the SHE are commonly estimated from the bulk spin Hall conductivity of the heavy metal, quantitative predictions of the second contribution are generally based on the Rashba model.\cite{PhysRevB.79.094422} Such simplified approaches are  unable to explain the sensitivity of the SOT to the substrate thickness or microscopic details of the interface.\cite{symmetry_spin_orbit_torques} Recently, a first-principles method to compute SOTs based on the general linear response formalism was developed,\cite{ibcsoit}
which allows us to fully take into account the fine details of the electronic structure crucial for determining the SOT in transition-metal multilayers with high accuracy.  The first application of this method to Co/Pt bilayers has shown very good agreement with experiment. More recently, the theory of the SOT arising in response to thermal gradients has  been also developed.\cite{Freimuth:151412} This allows us to access this spincaloric effect from {\it ab initio}.

From the practical point of view perpendicular magnetic anisotropy is desirable for various applications in~e.g.~data storage.\cite{179719} Many ferromagnetic multilayers of $3d$ transition metals with heavy
transition metals are known to exhibit very large magnetocrystalline anisotropy energy (MAE) that favors out-of-plane magnetization,\cite{PhysRevB.63.144409} and among materials
of this type Co/Pt bilayers are most studied experimentally with respect to SOT.\cite{PhysRevLett.109.096602,Miron:155001,MihaiMiron:155006} However, the large lattice mismatch between Co and Pt results in a rather poor quality of the interface in these systems. On the other hand, for disentangling various 
contributions to the SOT and for comparison between theoretical results with experiments 
in this kind of systems the high quality of the interface is of utter importance. In this work, we study from first principles the SOT in L1$_{0}$-FePt/Pt bilayers, which have a large out-of-plane
MAE,\cite{:/content/aip/journal/apl/100/14/10.1063/1.3700746} and which can be grown epitaxially
thus exhibiting high interfacial crystallinity.\cite{:/content/aip/journal/jap/109/7/10.1063/1.3556782,:/content/aip/journal/apl/90/13/10.1063/1.2717516}

The goal of this paper is two-fold. First, we compute and analyze different contributions to the SOT in L1$_0$-FePt/Pt thin films as a function of Pt thickness.\cite{ibcsoit} We analyze the energy dependence of the SOT and relate it to the energy dependence of the bulk spin Hall effect in Pt.
Secondly, taking FePt/Pt bilayers as an example, we present {\it ab initio} calculations of thermal SOT (T-SOT),~i.e.,~the SOT which is driven by a temperature gradient rather than an electric field. We briefly outline the ways of how the T-SOT can be enhanced. 
Finally we show that the energy dependence and magnitude of the even T-SOT can be estimated from the spin Nernst effect in bulk Pt.

\section{Formalism}
We investigate the SOT in our system using expressions obtained from the Kubo linear response formalism, and evaluated from the density functional theory. Within linear response the torque $\vn{T}$ exerted on the ferromagnetic magnetization when an electric field $\vn{E}$ is applied is given by $\vn{T} =\bf{t}\vn{E}$. The torkance tensor $\vn{t}$ has three contributions: \cite{ibcsoit} 
\begin{equation}\label{eq_torque}
\begin{aligned}
{\rm t}^{\rm I(a)\phantom{I}}_{ij}\!\!\!\!=-\frac{e}{h}\int_{-\infty}^{\infty}&
d\mathcal{E}
\frac{d f(\mathcal{E})}{d \mathcal{E}}
\phantom{\Re}
{\rm Tr}
\langle\mathcal{T}_{i}
G^{\rm R}(\mathcal{E})
v_{j}
G^{\rm A}(\mathcal{E})
\rangle,
\\
{\rm t}^{\rm I(b)\phantom{I}}_{ij}\!\!\!\!=\phantom{-}\frac{e}{h}\int_{-\infty}^{\infty}
&d\mathcal{E}\frac{d f(\mathcal{E})}{d \mathcal{E}}
{\Re}
{\rm Tr}
\langle\mathcal{T}_{i}
G^{\rm R}(\mathcal{E})
v_{j}
G^{\rm R}(\mathcal{E})
\rangle,\phantom{(1)}
\\
{\rm t}^{\rm II\phantom{(a)}}_{ij}\!\!\!\!=\phantom{-}\frac{e}{h}\int_{-\infty}^{\infty}
&d\mathcal{E} f(\mathcal{E})
\quad\!\!
{\Re}{\rm Tr}\langle
\mathcal{T}_{i}G^{\rm R}(\mathcal{E})v_{j}
\frac{dG^{\rm R}(\mathcal{E})}{d\mathcal{E}}\\
 &\quad\quad\quad\quad\quad\,-
\mathcal{T}_{i}\frac{dG^{\rm R}(\mathcal{E})}{d\mathcal{E}}v_{j}G^{\rm R}(\mathcal{E})
\rangle,
\end{aligned}\raisetag{4\baselineskip}
\end{equation}
with $G^{\rm R}(\mathcal{E})$ and $G^{\rm A}(\mathcal{E})$ as retarded and advanced Green functions, $v_{j}$ as the $j$th cartesian component of the velocity operator, $\mathcal{T}_{i}$ as the $i$th cartesian component of the torque operator, $f(\mathcal{E})$ as the Fermi distribution function and $e>0$ as the elementary positive charge. The torque operator is given by $\vht{\mathcal{T}}=\vn{m}\times\vn{B}^{\rm xc}$ where $\vn{m}$ and $\vn{B}^{\rm xc}$ are the spin magnetic moment operator and the exchange field, respectively. We model the influence of disorder in the system by a constant effective band broadening. Within this model 
the retarded and advanced Green functions are given by 
$G^{\rm R}(\mathcal{E})=\hbar[\mathcal{E}-H+i\Gamma]^{-1}$ and
$G^{\rm A}(\mathcal{E})=\hbar[\mathcal{E}-H-i\Gamma]^{-1}$, with parameter
$\Gamma$ characterizing the disorder strength. In this work we focus mainly on results
obtained for $\Gamma=25$\,meV, which corresponds to experiments performed at
room temperature, if the main source of disorder in the system is due to phonons.
In bulk metallic systems the diagonal and transverse conductivities at room temperature are usually reasonably well reproduced with this choice of $\Gamma$.\cite{PhysRevB.77.165117}

We decompose the torkance tensor $\vn{t}$ into even and odd components with respect to the
direction of magnetization $\hat{\vn{M}}$: ${\rm t}_{ij}={\rm t}_{ij}^{\rm even}+{\rm t}_{ij}^{\rm odd}$. It is very insightful to consider the limit $\Gamma \to 0$. In this so-called clean limit the even and odd components of the torkance tensor acquire qualitatively different forms:
\begin{equation}\label{eq_torque_even}
\mathrm{t}^{\rm even}_{ij}
=
\frac{2e}{\mathcal{N}}
\hat{\vn{e}}_{i} \cdot
\sum_{\vn{k},n}
f(\epsilon_{\vn{k}n})
\left[ 
\hat{\mathbf{M}}\times {\Im}
\Braket{ 
\frac{\partial u_{\vn{k}n}}{\partial\hat{\mathbf{M}}}| 
\frac{\partial u_{\vn{k}n}}{\partial k_{j}}   
}
\right],
\end{equation}
and
\begin{equation}\label{eq_torque_odd}
\mathrm{t}^{\rm odd}_{ij}
=-\frac{e\hbar}{2\Gamma\mathcal{N}}
\sum_{\vn{k}n}\langle\psi_{\vn{k}n}|\mathcal{T}_{i}|\psi_{\vn{k}n}\rangle
\langle\psi_{\vn{k}n}|v_{j}|\psi_{\vn{k}n}\rangle
\frac{\partial f (\epsilon_{\vn{k}n})}{\partial \mathcal{E}},
\end{equation}
where $\mathbf{k}$ is the Bloch vector in the Brillouin 
zone with an overall number $\mathcal{N}$,
$n$ runs over all bands, $\epsilon_{\vn{k}n}$ are the eigenenergies of the system,
$\psi_{\vn{k}n}$ and $u_{\vn{k}n}$ are the Bloch states and their lattice-periodic parts, respectively, and $\hat{\vn{e}}_{i}$ is the unit vector along the $i$th cartesian direction.   As discussed in other works by the authors,\cite{ibcsoit} the even torkance has the form of a Berry curvature and it is independent of $\Gamma$ in the limit of $\Gamma \to 0$. It constitutes the intrinsic contribution to the torkance, and it is analogous to the intrinsic anomalous or spin Hall effects. The 
odd part of the torkance, on the other hand, diverges like $1/\Gamma$ in the limit of small $\Gamma$,~i.e.,~it is proportional to the quasi-particle lifetime in analogy to the Rashba torque\cite{V.M._Edelstein_1990} or the diagonal electrical conductivity,\cite{PhysRevB.77.165117}, and it is thus dependent on the scattering mechanisms present in the system. 

Similarly to the spin Hall or anomalous Hall conductivities, the torkance tensor gives the SOT arising from an applied electric field,~i.e., it corresponds to a situation of a torque driven by a mechanical force. A torque can also be induced by a temperature gradient 
$\nabla \it{T}$,~i.e.,~it can also originate from statistical forces. Within linear response this thermal torque reads:
\begin{equation} 
\vn{T}=-\boldsymbol{\beta}\,\nabla T,
\end{equation} 
where $\boldsymbol{\beta}$ is the {\it thermal torkance}.
In analogy to the torkance driven by electrical currents, we decompose the thermal torkance into even and odd
components with respect to the magnetization direction. The intrinsic even part of the thermal torkance 
is analogous to the intrinsic 
anomalous Nernst\cite{PhysRevLett.97.026603,PhysRevB.87.060406}
and spin Nernst conductivities.\cite{SNE1,SNE2,tauber,wimmer} Similar to the latter effects,
it can be shown that the thermal torkance $\beta$ can be computed directly from its mechanical counterpart employing the Mott relation:\cite{Freimuth:151412}
\begin{equation}\label{eq_mott}
\beta_{ij}(T)=-\frac{1}{e}\int d\mathcal{E}\frac{\partial f(\mathcal{E},\mu,T )}{\partial\mu}
{\rm t}_{ij}(\mathcal{E})\frac{\mathcal{E}-\mu}{T}
\end{equation}
where ${\rm t}_{ij}(\mathcal{E})$ is the torkance tensor with Fermi energy set to $\mathcal{E}$ and $\mu$ is the chemical potential. 
In this work, we compute both electrical and thermal SOTs from the {\it ab initio} electronic struture of FePt/Pt bilayers according to Eqs.~(\ref{eq_torque}) and~(\ref{eq_mott}).

\section{Computational Details and basic properties}
In our study we considered 2 layers of L1$_0$-FePt oriented along [001]-axis and terminated with Fe atoms (Fe/Pt/Fe/Pt/Fe) deposited on the upper side of a Pt(001) film with the thickness of 6, 12 and 18 layers. 
The electronic structure of these L1$_0$-FePt/Pt(001) thin films was computed within the density functional theory using the Perdew, Burke, and Ernzerhof (PBE) functional and the full-potential linearized augmented-plane-wave method as implemented in the two-dimensional version of the code \texttt{FLEUR}.\cite{FLEUR}  DFT calculations were performed with 576 $k$-points in the two-dimensional Brillouin zone. The plane wave cutoff was set to 3.7$\,a_{0}^{-1}$ and the muffin-tin radii to 2.4\,$a_{0}$, where $a_{0}$ is the Bohr radius. The in-plane lattice constant of the films was set to the experimental lattice constant of fcc Pt (3.9265\,\AA). The out-of-plane relaxations of the atoms were performed until the forces were smaller
than $10^{-5}$\,Hartree/$a_{0}$, see Table~\ref{tab_atoms}.

\begin{table}[]
\caption{\label{tab_atoms}
Computational details for the thinnest film: interlayer distances ${\rm d_{z}}$ from one atomic layer to the next one (in units of ${\rm \AA}$); variation $\Delta=({\rm d_{z}}-{\rm d_{ref}})/{\rm d_{ref}}$ of the interlayer distances with ${\rm d_{ref}}={\rm d_{z}(Fe2)}$ for the first five atomic layers and ${\rm d_{ref}}={\rm d_{z}(Pt5)}$ for the other ones; spin magnetic moments $\mu_{{\rm at}}$ per atom (in units of $\mu_{B}$).
}
\begin{ruledtabular}
\begin{tabular}{cccc}
atomic layer &${\rm d_{z}}$ &$\Delta(\%)$ &$\mu_{{\rm at}}$\\
\hline
Fe1  &1.7896 &-3.9 &3.0804\\ 
Pt1  &1.8689 &0.4  &0.4032\\
Fe2  &1.8616 &0.0  &3.0213\\
Pt2  &1.8736 &0.6  &0.3829\\
Fe3  &1.8158 &-2.5 &3.0403\\
\hline
Pt3  &2.0998 &3.6  &0.2967\\
Pt4  &2.0393 &0.6  &0.0474\\
Pt5  &2.0271 &0.0  &0.0216\\
Pt6  &2.0193 &-0.4 &0.0093\\
Pt7  &1.9824 &-2.2 &0.0076\\
Pt8  &       &     &0.0071\\
\end{tabular}
\end{ruledtabular}
\end{table}

For magnetization out-of-plane the computed spin moments of Fe atoms range between 3.02$\,\mu_B$ and 3.08$\,\mu_B$ depending on the thickness and position of the Fe atom with respect to the interface with the Pt substrate.
The largest spin moment of the Pt atoms is about 0.4$\,\mu_B$ in the FePt overlayer, while the largest spin moment among the substrate atoms is 0.3$\,\mu_B$ for the Pt atom closest to the interface. Spin moments then rapidly decay when going further in the substrate (see also Table~\ref{tab_atoms}). For the thinnest film we have also computed the value of the magnetocrystalline anisotropy energy and found it to be 1.2\,meV per Fe atom favoring the
out-of-plane magnetization, while the anisotropy within the plane was one order of magnitude smaller.
 
For computing the SOTs we employed the Wannier interpolation technique. We constructed 18 maximally localized Wannier functions (MLWFs) per atom from Bloch functions on an 8$\times$8 $k$-point mesh using the wannier90 program.\cite{WannierPaper,Mostofi2008685} The number of bands used to disentangle the subspace of the MLWFs was chosen such that for each film the ratio of the number of bands to the number of MLWFs 
was approximately equal to 1.4. This allows a very precise interpolation of the electronic structure up to 5\,eV above the Fermi energy. The torkances were computed on a 2048$\times$2048 $k$-point mesh, except for the case of $\Gamma$ well below 25\,meV, where a 4096$\times$4096 $k$-point mesh was used. For magnetization out-of-plane, which is the case
considered here, the only non-vanishing independent components of the torkance tensor are 
${\rm t}^{\rm even}_{yx}$ and ${\rm t}^{\rm odd}_{xx}$, with the convention that the $z$ axis points out-of-plane, while the $x$ and $y$ axes coincide with the [100] and [010] in-plane directions.

\section{Results}
\subsection{Spin-orbit torques driven by electrical currents}

\begin{figure}[t!]
\centering
\includegraphics*[width=4.1cm]{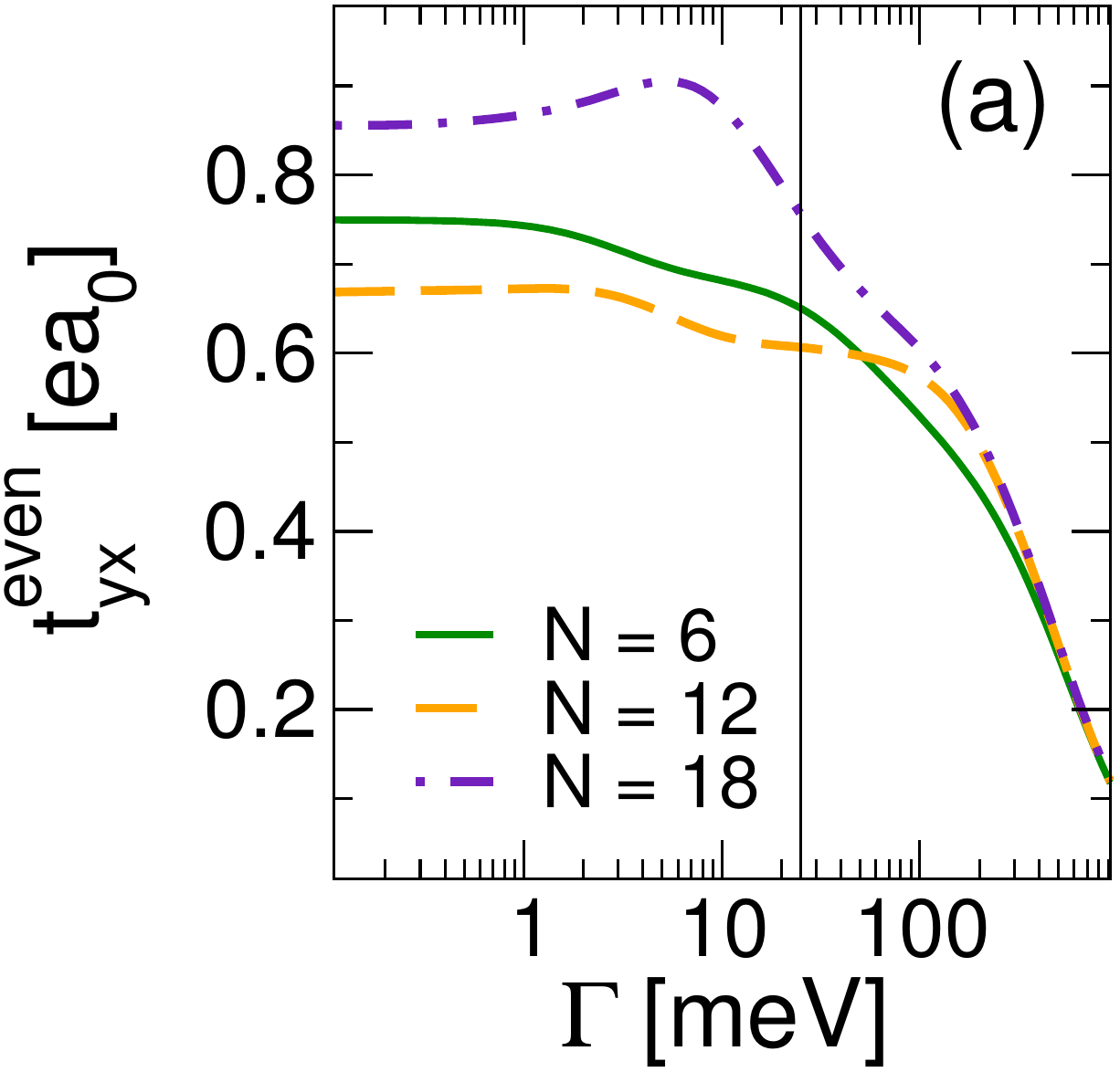}
\hspace{0.1cm}
\includegraphics*[width=4.1cm]{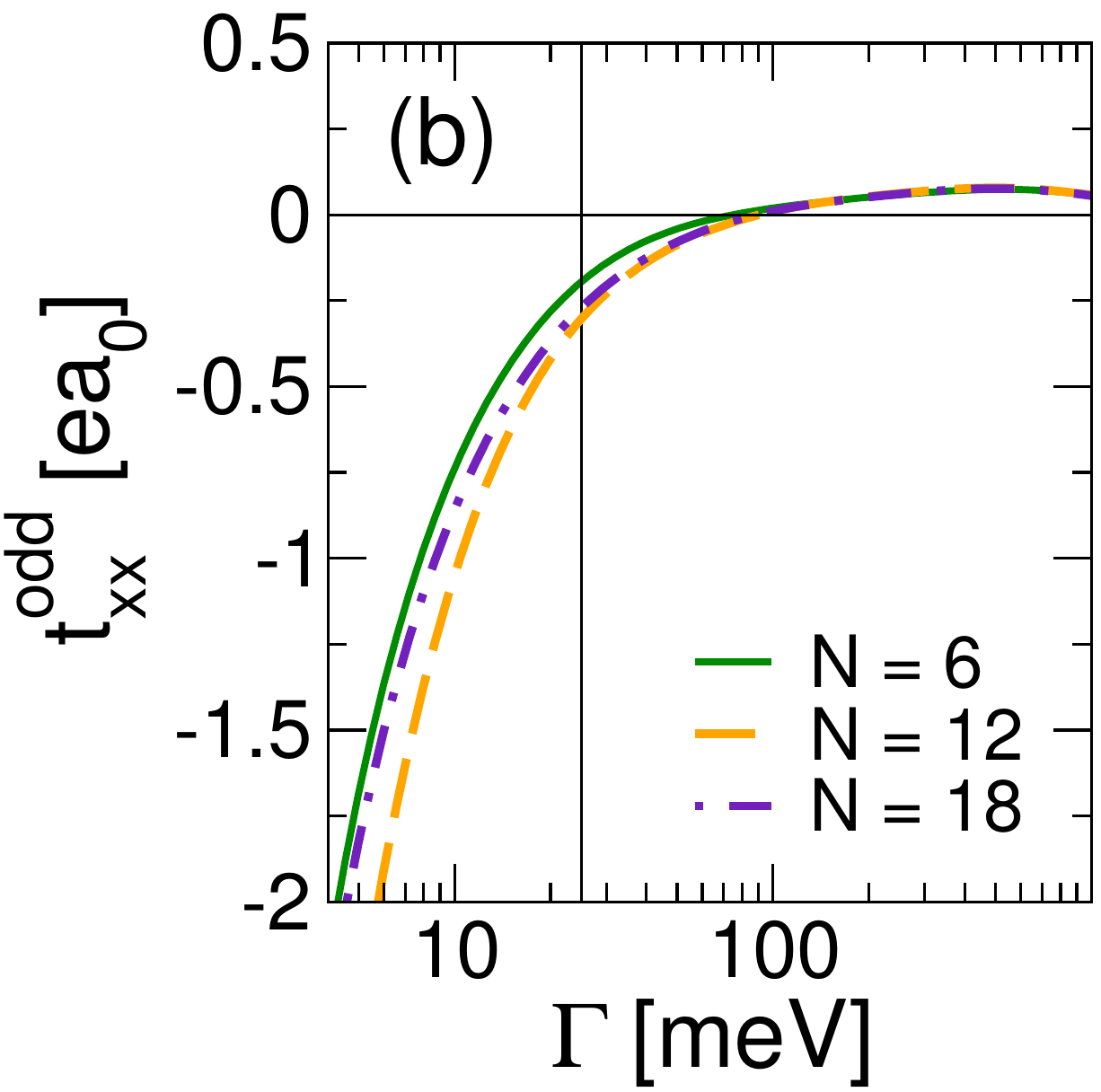}
\caption{\label{fig_torque_gamma}
a) Even torkance ${\rm t}^{\rm even}_{yx}$ and b) odd torkance ${\rm t}^{\rm odd}_{xx}$ in L1$_0$-FePt$^{2}$/Pt$^{\mathrm{N}}$ for N = 6 (green solid), 12 (orange dashed) and 18 (blue dot-dashed), as a function of the disorder strength $\Gamma$. Solid vertical lines correspond to the value of 
$\Gamma=25$~meV.
}
\end{figure}

We first compute the even and odd torkance as a function of the disorder strength $\Gamma$ and thickness of the Pt substrate using the expressions from the previous section. The results
of these calculations are presented in Fig.~\ref{fig_torque_gamma} and summarized in 
Table~\ref{tab_effective_fields} for the band broadening of $\Gamma=25$\,meV$\,\approx k_BT_0$, which mimicks the effect of the room temperature $T_0$. 
At small $\Gamma$ the even torkance ${\rm t}^{\rm even}_{yx}$ is given by its clean limit Berry curvature
value which lies in the range of  0.65 to 0.85$\,ea_0$ depending on the substrate thickness, 
and the deviation of ${\rm t}^{\rm even}_{yx}(\Gamma)$ from these values becomes significant only for band broadening larger than
100~meV. In the latter case the values of ${\rm t}^{\rm even}_{yx}$ for different numbers of Pt layers 
are almost identical to each other, meaning that the fine difference in the electronic structure
of the films is washed out by the broadening of this magnitude. 
At $\Gamma=25$\,meV the even torkance is still relatively close to the Berry curvature values,
see also Table~\ref{tab_effective_fields}, and the variation in ${\rm t}^{\rm even}_{yx}$ caused by Pt thickness
is of the order of 15\%. For this broadening the values of ${\rm t}^{\rm even}_{yx}$ for our system are
rather close to those of Co$^3$/Pt$^{10}$(111) bilayers, as computed in 
Ref.~\onlinecite{ibcsoit}, which lie in the range of 0.53 to 0.62$\,ea_0$ depending 
on the capping. 

As for the odd torkance, for broadenings below 10\,meV its magnitude
is larger than that of the even torkance, while ${\rm t}^{\rm odd}_{xx}$ rapidly decays with $\Gamma$
and changes sign in the vicinity of $\Gamma\approx 80$\,meV, where the difference in 
${\rm t}^{\rm odd}_{xx}$ for films of different thickness is almost negligible. Overall, the 
characteristic $1/\Gamma$-behavior is clearly visible for small $\Gamma$. At room temperature
the odd torkance is negative and it is roughly twice smaller in magnitude than the 
corresponding even torkance. The fact that ${\rm t}^{\rm odd}_{xx}$ is close to the point of changing the sign for $\Gamma=25$\,meV makes it also more sensitive to the Pt thickness, which 
otherwise does not have a pronounced effect on the odd torkance 
(see also Table~\ref{tab_effective_fields}).

For comparing to experiments it is useful to represent the
computed torkances in terms of the effective magnetic fields at a given current 
density, and in Table~\ref{tab_effective_fields} we present the corresponding values
of T$^{\rm even}_{y}/\mu_{s}$ and T$^{\rm odd}_{y}/\mu_{s}$ for an 
electric field $E_x$ of 360\,V/cm,
where $\mu_{s}$ stands for the total spin moment in the unit cell containing three Fe atoms with the value
of about 10.1$\,\mu_B$ for all thicknesses and magnetization out-of-plane.
The value of the electric field chosen to compute the effective magnetic fields corresponds 
to the current density $j\approx 10^{7}$A/cm$^{2}$, if one estimates the order of the resistivity of our L1$_0$-FePt/Pt thin films by the experimentally measured room temperature resistivity of the Pt/Co/AlO$_{x}$ system.\cite{symmetry_spin_orbit_torques} The values of the even effective magnetic fields of the order of 2.0\,mT are generally consistent with those computed for Co/Pt bilayers,\cite{ibcsoit} taking into account that the value of $\mu_s$ in the latter case is smaller by 
about 30\% than that in FePt/Pt bilayers that we study here. The magnitude of T$^{\rm odd}_{x}/\mu_{s}$ in FePt/Pt bilayers is, on the other hand, significantly smaller than the magnitude of T$^{\rm even}_{y}/\mu_{s}$, see Table~\ref{tab_effective_fields}.

\begin{table}[]
\caption{\label{tab_effective_fields}
Even and odd torkances $\rm t$ computed at $\Gamma$ = 25\,meV (in units of $ea_{0}$);
even (T$^{\rm even}_{y}/\mu_{s}$) and odd (T$^{\rm odd}_{x}/\mu_{s}$)
effective magnetic fields (in units of mT) for an applied electric field $E_{x}=360\,$V/cm; even and odd thermal torkances $\beta$ (in units of
 $\mu e$V$\cdot a_{0}\cdot$K$^{-1}$);
$|\nabla T|^0$ (in units of K/nm) is the temperature gradient required to reproduce the total effective magnetic field $\sqrt{({\rm T}^{\rm odd}_{x})^{2}+({\rm T}^{\rm even}_{y})^{2}}/\mu_{s}$.
}
\begin{ruledtabular}
\begin{tabular}{cccc}
&FePt$^{2}$/Pt$^{6}$ &FePt$^{2}$/Pt$^{12}$ &FePt$^{2}$/Pt$^{18}$\\
\hline
${\rm t}^{\rm even}_{yx}$ &+0.65 &+0.61 &+0.75 \\
${\rm t}^{\rm odd}_{xx}$ &$-$0.19 &$-$0.30 &$-$0.27 \\
T$^{\rm even}_{y}/\mu_{s}$ &+2.1 &+2.0 &+2.4\\
T$^{\rm odd}_{x}/\mu_{s}$ &$-$0.6 &$-$1.0 &$-$0.9\\
$\beta^{\rm even}_{yx}$ &$-$10.6 &$-$15.3 &$-$14.5 \\
$\beta^{\rm odd}_{xx}$ &$-$4.8 &+0.7 &$-$2.5 \\
$|\nabla T|^0$ &+2.1 &+1.6 &+1.7\\
\end{tabular}
\end{ruledtabular}
\end{table}

It is tempting to compare the computed even SOT to the ``hypothetical" torque ${\rm T}_{y}$ that is exerted on the magnetization if the spin current density $j^{y}_{z}$ accross the interface between Pt substrate and L1$_0$-FePt overlayer is given by the spin Hall conductivity of bulk fcc Pt. In that case the current density $j^{y}_{z}$ generated by the spin Hall effect is given by the relation $j^{y}_{z}=\sigma^{y}_{zx}E_{x}$ when an electric field $E_x$ is applied to the system. In the latter expression $\sigma^{y}_{zx}$ stands for the corresponding component of the spin Hall conductivity (SHC) tensor of bulk Pt. Under the assumption that the whole of the bulk spin Hall current is transferred to the magnetization,~i.e.,~that ${\rm T}_{y}=S j^{y}_{z}$, where the spin polarization of the spin current is along the $y$-axis and  $S$ = 7.712\,\AA$^{2}$ is the in-plane area of the unit cell, this model yields a simple expression for the even torkance:

\begin{equation}\label{eq_torque_model}
{\rm t}^{\rm SHE}_{yx}=S\vn{\sigma}^{y}_{zx}.
\end{equation}

\begin{figure}[t!]
\centering
\includegraphics*[width=8.5cm]{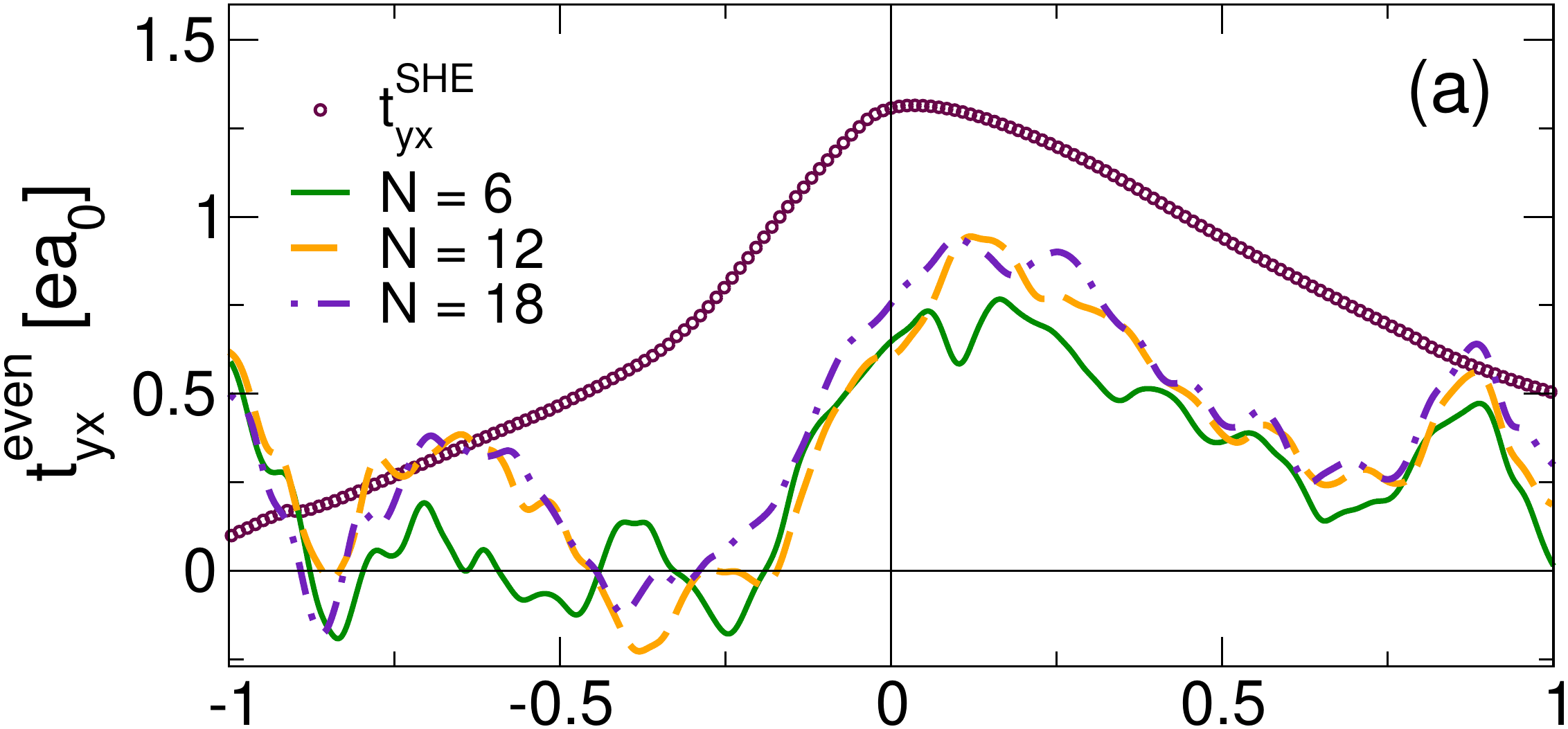}
\includegraphics*[width=8.5cm]{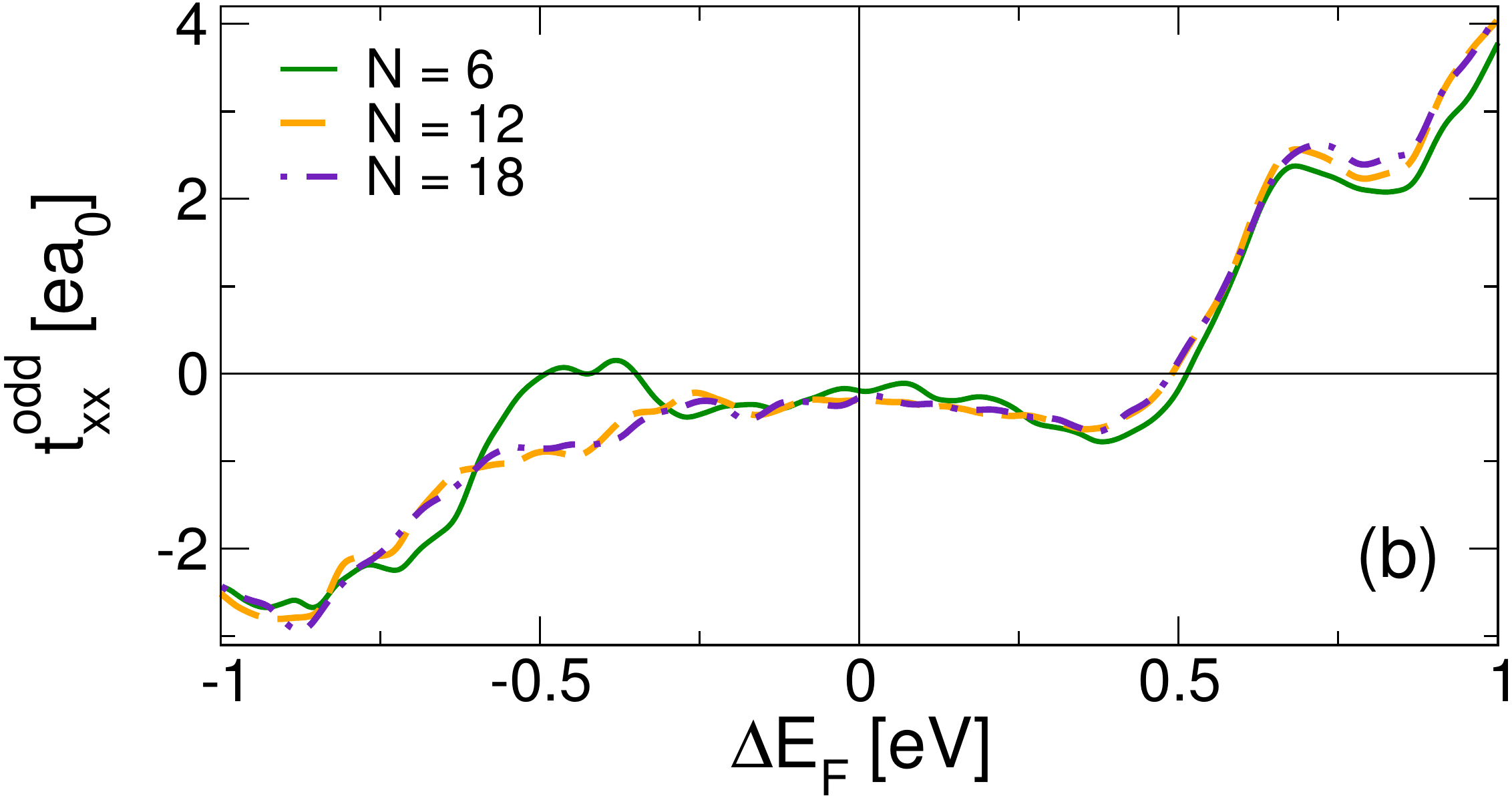}
\caption{\label{fig_torque}
(a) Even torkance ${\rm t}^{\rm even}_{yx}$ and (b) odd torkance ${\rm t}^{\rm odd}_{xx}$  as a function of
the Fermi energy (with respect to the true Fermi energy ${\rm E_{F}}\approx -4.33$ eV for all three thicknesses) at $\Gamma$ = 25\,meV in L1$_0$-FePt$^{2}$/Pt$^{\mathrm{N}}$ films for N = 6 (green solid), 12 (orange dashed) and 18 (blue dot-dashed). The line of circles in the upper figure corresponds to the even torkance ${\rm t}^{\rm SHE}_{yx}$ estimated from the spin Hall conductivity of bulk fcc Pt,~Eq.~\eqref{eq_torque_model}.  
}
\end{figure}

In Fig.~\ref{fig_torque} we plot the even spin Hall torkance ${\rm t}^{\rm SHE}_{yx}$
in comparison to the even torkance ${\rm t}_{yx}$ computed at $\Gamma=25$\,meV as a function of the Fermi energy in our
system. For estimating ${\rm t}^{\rm SHE}_{yx}$ we used the intrinsic SHC in bulk fcc Pt (note that our calculations show that
the influence of 
the band smearing of the order of 25\,meV on the clean limit SHC is negligible). At the true Fermi energy, the SHC of fcc Pt is found to be
2184\,$(\hbar/e)$S/cm. As apparent from Fig.~\ref{fig_torque}, in the interval
of energies of $[-0.1, +0.5]$\,eV with respect to the true Fermi energy, the even SOT can be 
approximated with the expression ${\rm t}^{\rm even}_{yx}=\xi \,{\rm t}^{\rm SHE}_{yx}$, where the so-called {\it
SHE-to-SOT efficiency} $\xi$\cite{Inv_SOT} smoothly varies with energy in the range of 
$0.5<\xi<0.7$ and moderately depends on the Pt thickness. As a result, in this energy range  the qualitative behavior of ${\rm t}^{\rm even}_{yx}$ quite closely resembles that of ${\rm t}^{\rm SHE}_{yx}$.
In this energy region one could attribute the moderate energy and Pt thickness dependence  
of $\xi$ and its deviation from the ``ideal" value of 1.0 to the finite size effects
and details of the electronic structure which,~e.g., influence the magnitude of the spin current generated in the Pt substrate,
as well as its $z$-distribution inside the slab and transmission properties of the interface.\cite{Inv_SOT} 
We note that the range of values of $\xi$ for energies between $-0.1$ and $0.5$\,eV
is rather close to that computed in the presence of disorder for Co/Pt bilayers. For the latter system  it was shown that ${\rm t}^{\rm even}_{yx}$ arises
mainly due to the spin current which originates from the SHE inside the Pt substrate.\cite{ibcsoit}
On the other hand, away from this energy range, ${\rm t}^{\rm even}_{yx}$ in FePt/Pt can 
differ from ${\rm t}^{\rm SHE}_{yx}$
by an order of magnitude and even in sign (e.g. around ${\rm E_{\rm F}}=-0.35$ and $-$0.8\,eV),
which signifies that the application of simplified models of the kind of Eq.~\eqref{eq_torque_model} has to be done with extreme caution. 

\begin{figure}[t!]
\centering
\includegraphics*[width=8.5cm]{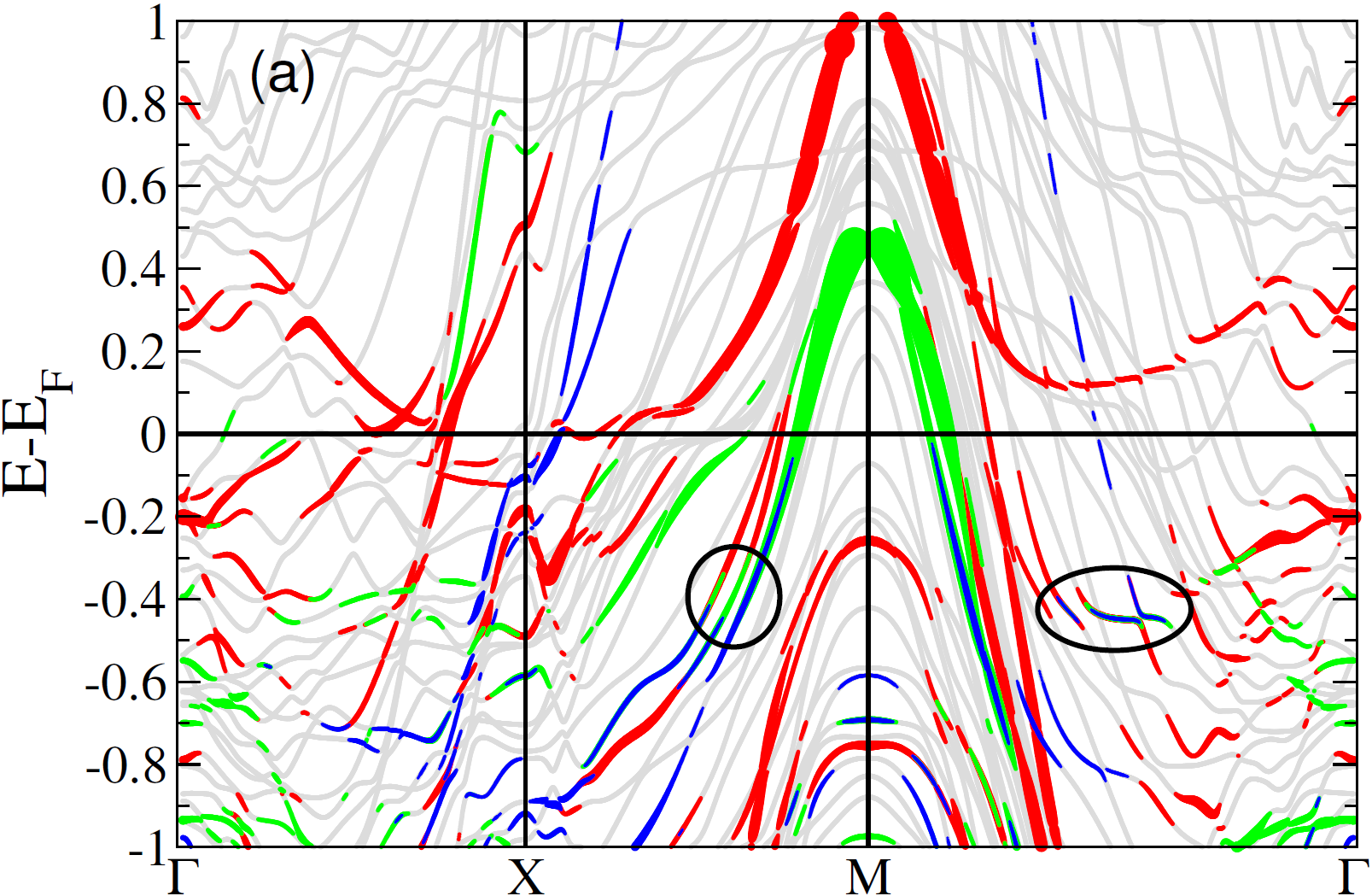}
\hspace{0.2cm}
\includegraphics*[width=8.5cm]{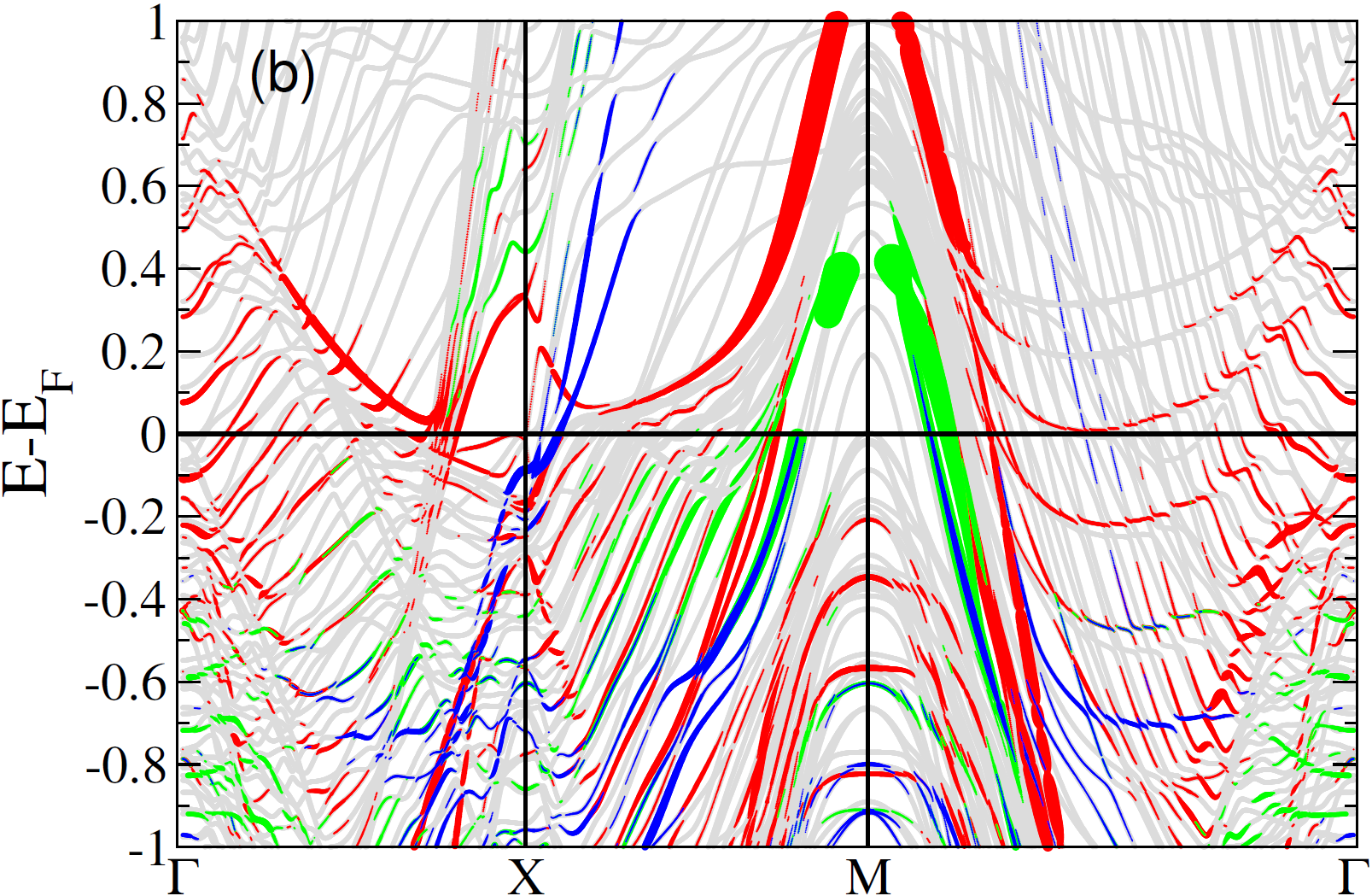}
\caption{\label{fig_bands}
Band structures of (a) L1$_0$-FePt$^{2}$/Pt$^{6}$  and (b) L1$_0$-FePt$^{2}$/Pt$^{18}$  thin films along high symmetry lines. States with large portion of the wavefunction on specific atoms are marked by red (Pt atoms at the bottom of the slab), green (Pt substrate atoms closest to the FePt/Pt interface) and blue (Fe atoms closest to the FePt/Pt interface). The criteria for a state to be marked is that a) more than 9.6\% for Pt-atoms and 7.7\% for Fe-atoms of the charge of the state are localized inside a corresponding atom. For b)  these values constitute 4.5\% for Pt-atoms and 3.6\% for Fe-atoms, owing to the twice larger thickness. The radius of the dots is proportional to the weight of the wavefunction inside a corresponding atom. All states are marked by grey dots in background.
}
\end{figure}

Fig.~\ref{fig_torque} shows that while the even torkance as a function of energy 
reaches its maximal values around the true Fermi energy of FePt/Pt, the values of the 
odd torkance are small around the true ${\rm E_{\rm F}}$, and they become very large away from
it. As far as the thickness dependence of both ${\rm t}^{\rm even}_{yx}$ and ${\rm t}^{\rm odd}_{xx}$
is concerned, significant deviations between the torkances for ${\rm N}=6$ and larger
thicknesses are visible only in the energy interval of about $-0.7$\,eV to $-0.3$\,eV. 
The difference in the torkances for ${\rm N}=12$ and ${\rm N}=18$ is, on the other
hand, smaller. Among the two, the thickness dependence is more pronounced for the
odd torkance with the difference reaching as much as $1\,ea_0$ between ${\rm t}^{\rm odd}_{xx}$ 
for thin and thick films, while for ${\rm t}^{\rm even}_{yx}$ this difference is much smaller. 
In Fig.~\ref{fig_bands} we present the bandstructures of the slabs with 6 and 18 layers of Pt in the substrate.
The two bandstructures look very similar with the only obvious difference lying in the increased number of bands for the
thicker substrate.
However, in the energy interval of interest a relatively large hybridization of the states which have larger weight at 
the bottom layer of the Pt substrate with 
the states which exhibit larger weight at the interface between L1$_0$-FePt and Pt is clearly
visible for the L1$_0$-FePt$^{2}$/Pt$^{6}$ film (see black circles in Fig.~\ref{fig_bands}), while this hybridization is almost absent for the L1$_0$-FePt$^{2}$/Pt$^{18}$ film. Thus, we speculate that the cross-talk between the free surface of the
Pt substrate and the interface with FePt, which are almost decoupled for large Pt thicknesses, and quite pronounced for the 6-layer film, could lead to significant differences in the SOTs of thin and thick FePt/Pt bilayers.

\subsection{Thermal spin-orbit torques}

\begin{figure}[t!]
\centering
\includegraphics*[width=8.5cm]{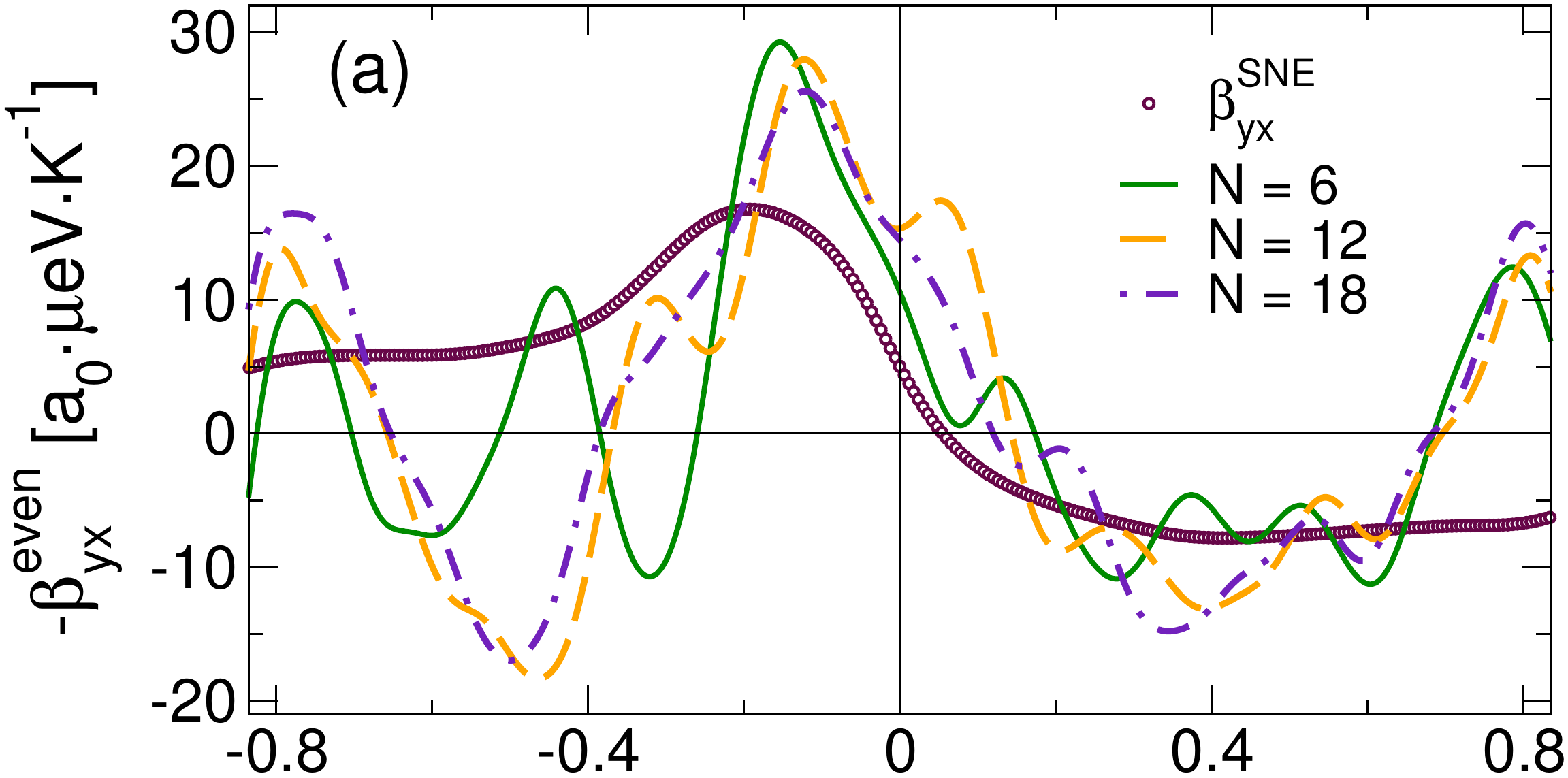}
\includegraphics*[width=8.5cm]{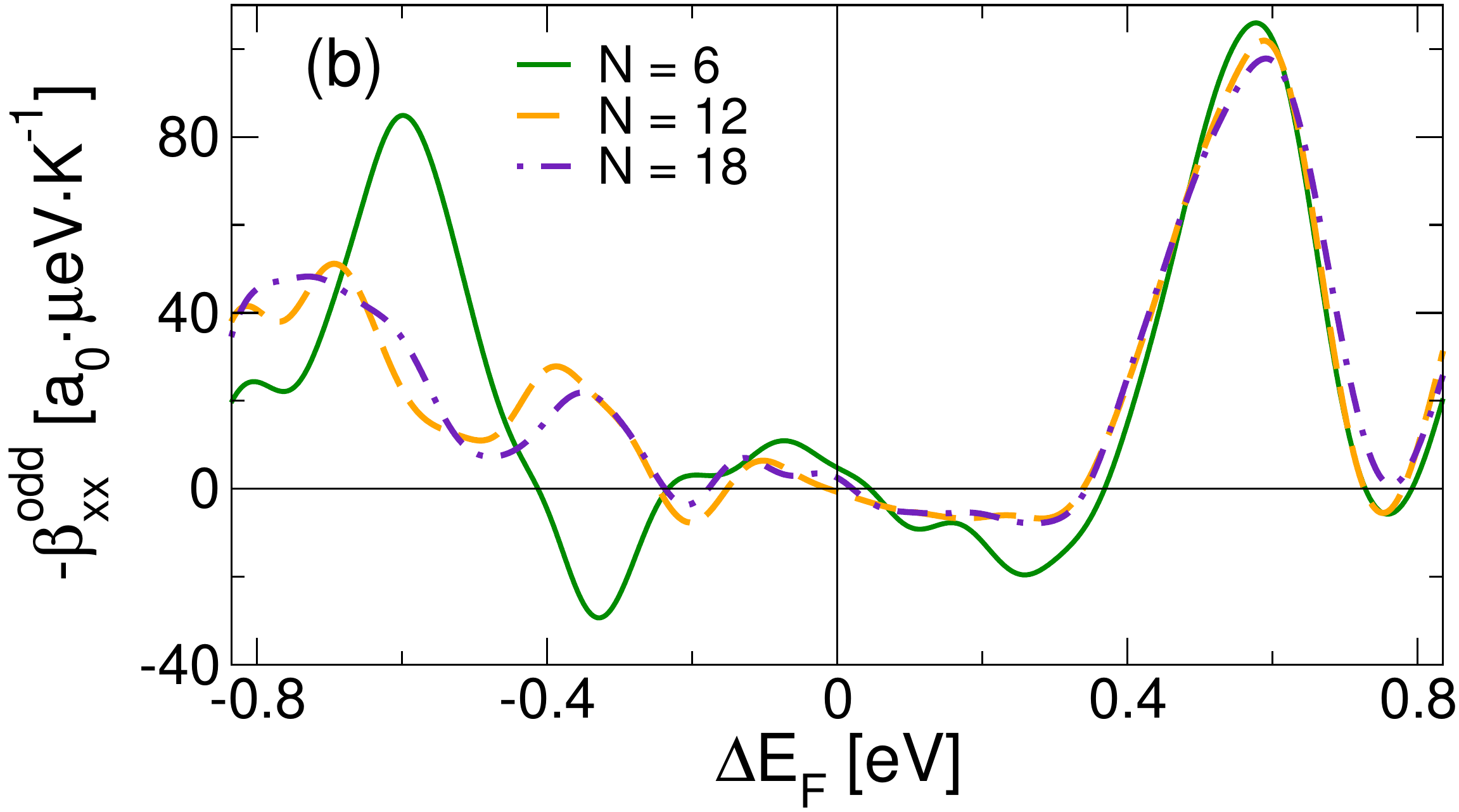}
\caption{\label{fig_th_torque}
a) Even thermal torkance $\beta^{\rm even}_{yx}$ and b) odd thermal torkance $\beta^{\rm odd}_{xx}$ are calculated for $T$ = 300\,K using Eq.~\eqref{eq_mott}, based respectively on ${\rm t}^{\rm even}_{yx}$ and ${\rm t}^{\rm odd}_{xx}$ from Fig.~\ref{fig_torque}. The line of circles in the upper figure corresponds to the even thermal torkance ${\rm \beta}^{\rm SNE}_{yx}$ estimated from the spin Nernst conductivity of bulk fcc Pt,~Eq.~\eqref{eq_thtorque_model}.
}
\end{figure}

We compute the thermal spin-orbit torques (T-SOTs) in our system according to 
Eq.~\eqref{eq_mott} at temperature ${\it T}=300$\,K using as input the energy dependence
of the even and odd torkances computed at $\Gamma=25$\,meV and presented in 
Fig.~\ref{fig_torque}. The energy dependence of the even and odd thermal torkances
$\beta^{\rm even}_{yx}$ and $\beta^{\rm odd}_{xx}$ of L1$_0$-FePt/Pt thin films 
at room temperature is shown in Fig.~\ref{fig_th_torque}, and their values at the
Fermi energy are summarized in Table~\ref{tab_effective_fields}. 
 
By direct inspection, it is easy to see that the trend of $\beta^{\rm even}_{yx}$ and $\beta^{\rm odd}_{xx}$ with energy can be directly related to the corresponding behavior of ${\rm t}^{\rm even}_{yx}$ and ${\rm t}^{\rm odd}_{xx}$. This follows from the observation that in the limit of zero 
temperature $T$ in Eq.~\eqref{eq_mott} the thermal torkance $\beta$ is proportional to the energy
derivative of the torkance $\rm t$ at the corresponding energy. Indeed, by comparing
the curves in Figs.~\ref{fig_th_torque} and~\ref{fig_torque}, we can see that in most of the
cases the zeros of the thermal torkance correspond to the local extrema of the electrical
torkance, while the maxima in the former correspond to the regions of largest slope of
the latter. It is thus not surprising that the largest values of $\beta^{\rm even}_{yx}$ of 
the order of tens of  $\mu e$V$\cdot a_{0}\cdot$K$^{-1}$ are achieved around the Fermi energy,
while the magnitude of $\beta^{\rm odd}_{xx}$ is maximal away from the Fermi energy,
reaching as much as 100\,$\mu e$V$\cdot a_{0}\cdot$K$^{-1}$ there. Clearly visible
in Fig.~\ref{fig_th_torque} is a much more pronounced dependence of the thermal torkances on the
Pt thickness than in the case of the electrical torkances. The thermal torkances for 6 and 12/18
layers of Pt substrate differ in sign over wide patches in energy around $-0.4$\,eV and the
difference between thermal torkances for 12 and 18 layers becomes more pronounced. At the true Fermi energy,
$\beta^{\rm odd}_{xx}$ exhibits a change of sign when changing the Pt thickness, see Table~\ref{tab_effective_fields}.

It is known that in paramagnetic metals, in particular Pt, an applied temperature gradient will
result in transverse spin current, analogous to the spin Hall current which is generated by an electric
field. The respective phenomenon is called the spin Nernst effect (SNE),\cite{SNE1,SNE2} and its
magnitude is characterized by the spin Nerst conductivity (SNC) $\alpha$. Keeping in mind the geometry of our system, the relationship between a temperature gradient applied 
along the $x$ axis and the spin current density with spin-polarization along the $y$ axis which propagates
along the $z$ axis, reads: 
\begin{equation}
 j^{y}_{z}=-\alpha^{y}_{zx}\nabla \it{T}_{x}.
\end{equation} 
As in the previous section, we will compare the magnitude of the ``pure" spin Nernst
torkance $\beta^{\rm SNE}_{yx}$ to the computed thermal torkance $\beta^{\rm even}_{yx}$,
assuming that the spin Nernst torkance arises from the full bulk spin Nernst current:
\begin{equation}\label{eq_thtorque_model}
\beta^{\rm SNE}_{yx}=S\vn{\alpha}^{y}_{zx}.
\end{equation}
To estimate the magnitude of the spin Nernst thermal torkance from {\it ab initio}, from the
energy dependence of the SHC presented in Fig.~\ref{fig_torque}, we evaluate the thermal intrinsic contribution to the SNC according to the Mott relation (at $T=300$\,K):\cite{tauber,wimmer}
\begin{equation}\label{eq_mott_sne}
\alpha^y_{zx}=-\frac{1}{e}\int d\mathcal{E}\frac{\partial f(\mathcal{E},\mu,\it{T}) }{\partial\mu}
\sigma^y_{zx}(\mathcal{E})\frac{\mathcal{E}-\mu}{\it{T}}.
\end{equation}
The spin Nernst thermal torkance $\beta^{\rm SNE}_{yx}$ computed using Eq.~\eqref{eq_thtorque_model} and Eq.~\eqref{eq_mott_sne} is presented in Fig.~\ref{fig_th_torque} as a function
of the position of the Fermi energy together with $\beta^{\rm even}_{yx}$. By comparing
the two torkances we can conclude that, as in the case of the electrical torkances,
the overall behavior of  $\beta^{\rm even}_{yx}$ with energy is in accordance with that
of $\beta^{\rm SNE}_{yx}$ in the window of energies between  $-0.2$\,eV and $+0.6$\,eV.
This hints at a clear correlation between the phenomenon of the T-SOT and the SNE at these
energies for our system. Owing to the essential energy dependence of the SHE-to-SOT efficiency
$\xi$, the {\it SNE-to-T-SOT efficiency} $\xi^T$, defined by relation $\beta^{\rm even}_{yx} = \xi^T
\beta^{\rm SNE}_{yx}$, deviates quite significantly from $\xi$ and ranges approximately 
between 0.5 and 1.5 in the energy interval [$-0.2$\,eV, $+0.6$\,eV], with the exception
of energies where the torkances change sign in between 0.0 and 0.2\,eV.

Since to the best of our knowledge the effect of T-SOT has not been observed so far,
it is important that we give an estimate of the T-SOT that can be achieved experimentally 
in our films. We therefore compute the temperature gradient $|\nabla T|^0$ that is required to reproduce the total effective magnetic field obtained with the value of current density $j \sim 10^{7}$\,A/cm$^{2}$, typical for experiments on such systems (Table~\ref{tab_effective_fields}). 
The value of $|\nabla T|^0$ of the order of 2\,K/nm which we obtain for our L1$_0$-FePt/Pt 
bilayers at their true Fermi energy turns out to be one order of magnitude larger than the one
which can be achieved experimentally in this type of systems.\cite{Wees} This means that although
the T-SOT in the system that we study here most probably cannot be used to switch the 
magnetization, we conclude that the fingerprints of the effect can be observed.

We are, moreover, confident that at the current level of experimental techniques the T-SOT 
can be made as large as the electrical SOT by proper electronic structure engineering, which can go along three different paths. (i) As apparent from Fig.~\ref{fig_th_torque}, for FePt/Pt bilayers 
the thermal torkances can be order of magnitude larger if the Fermi energy is set to $\sim$~0.6~eV above its true value - this corresponds roughly to using~e.g.~L1$_0$-(Fe$_{1-x}$Co$_{x}$)(Pt$_{1-x}$Au$_{x}$)/Pt$_{1-x}$Au$_{x}$ instead of FePt/Pt, with $x\sim$~0.6 if we assume a constant density of states of $\sim$~1~eV$^{-1}$ per atom for Fe$_{1-x}$Co$_{x}$Pt$_{1-x}$Au$_{x}$ and Pt$_{1-x}$Au$_{x}$. (ii) Exploiting the close correlation between the T-SOT and the SNE which
we found, one could consider using fcc Ir, Pd or Rh as substrates instead of fcc Pt, since the values of
the intrinsic SNCs for these metals which we computed constitute $-$8744 (Ir), $+$20804 (Pd), and $-$20779\,$(\hbar/e)$$\mu$A$\cdot$cm$^{-1}\cdot$K$^{-1}$ (Rh), which is respectively $+$1.04, $-$2.48 and $+$2.48 times larger than the value of the SNC of fcc Pt of $-$8383\,$(\hbar/e)$$\mu$A$\cdot$cm$^{-1}\cdot$K$^{-1}$. (iii)
Our calculations show that upon decreasing the disorder strength
$\Gamma$ the energy dependence of the odd and, particularly, even torkances exhibits strong deviations from the smooth behavior shown
above, acquiring sharp features and sign changes at the scale of tens of
meVs. This effect is due to the fine features in the electronic structure
of thin films, which get promoted as the band broadening is decreased.
Correspondingly, upon reducing the degree of disorder in the system (e.g.~by
lowering of the temperature or concentration of impurities) the magnitude
of the T-SOT, qualitatively proportional to the degree of raggedness
of the torkance as a function of energy, can be significantly enhanced,
as confirmed by our calculations.

\section{Conclusions}
Using expressions for the spin-orbit torkances derived from the Kubo linear response
formalism, we compute from first principles the values of the even and odd torkances in a system consisting of two layers of ferromagnetic L1$_0$-FePt deposited on an fcc Pt(001) substrate of various thicknesses. We predict that the magnitude of the SOTs lies in the range of values measured experimentally and computed theoretically for Co/Pt bilayers. For both even and 
odd torques we find a pronounced energy and thickness dependence. By comparing 
the even SOT to that purely given by the spin Hall effect in the Pt substrate we find that while around 
the Fermi energy the behavior of the two SOTs is very similar, they can differ in sign and order
of magnitude for wide regions of energy. Moreover, using the expressions that we derived recently
for the thermal SOT, driven by the temperature gradient rather than the electric field, we compute
the energy and thickness dependence of the thermal torkance in the system under consideration.
We were also able to establish a close connection between the T-SOT and the spin Nernst
effect. We predict that thermal gradients of the order of 2\,K/nm are necessary to exert the
same torque on the magnetization as that arising from typical current densities in this kind
of systems, which assures us that the T-SOT in FePt/Pt bilayers could be experimentally detected.
We further speculate that much larger T-SOTs can be achieved in other ferromagnetic 
transition-metal overlayers deposited on substrates which exhibit larger spin Nernst effect
than Pt.

We gratefully acknowledge computing time on the supercomputers JUQUEEN and JUROPA at 
J\"ulich Supercomputing Center as well as at the JARA-HPC cluster of RWTH Aachen, and funding under the HGF-YIG programme VH-NG-513 and SPP 1538 of DFG.
\bibliography{letter}

\end{document}